# Characterisation of the potential of frequency modulation and optical feedback locking for cavity-enhanced absorption spectroscopy

Vasili L. Kasyutich[*] and Markus W. Sigrist[**]

Institute for Quantum Electronics, Department of Physics, Swiss Federal Institute of Technology, Zurich, Switzerland

*kvasili@phys.ethz.ch

** sigristm@phys.ethz.ch

A combination of optical feedback self-locking of a continuous-wave distributed feedback diode laser to a V-shaped high finesse cavity, laser phase modulation at a frequency equal to the free spectral range of the V-cavity and detection of the transmitted laser beam at this high modulation frequency is described for possible application in cavity-enhanced absorption spectroscopy. In order to estimate an absorbance baseline noise of laser intensity and frequency modulated light triplet passed through the V-cavity in open air, a 1.5-cm long optical cell filled by $C_2H_2$ at low pressure was placed behind the cavity output mirror. The performance of the setup was evaluated from the experimental bandwidth normalised relative intensity noise on the cavity output and the frequency modulation absorption signals induced by $C_2H_2$ absorption in the 1.5-cm cell. From these data we estimate that the noise-equivalent absorption sensitivity of $2.1\times10^{-11}$ cm$^{-1}$Hz$^{-1/2}$ by a factor of 11.7 above a shot-noise limit can be achieved for $C_2H_2$ absorption spectra extracted from the heterodyne beat signals recorded at the transmission maxima intensity peaks of the successive TEM$_{00}$ resonances.

**Introduction**

The highest sensitivities for absorption measurements close to a shot noise limit were demonstrated in noise-immune cavity-enhanced optical heterodyne molecular spectroscopy (NICE-OHMS) [1]. In NICE-OHMS the laser frequency is modulated at high modulation frequency (>100 MHz) equal to a multiple of a free spectral range (FSR) of a high finesse cavity, whilst absorption and dispersion signals can be extracted from the transmitted optical field by means of a lock-in amplifier at high modulation frequency. However, routine demonstrations of NICE-OHMS absorption sensitivities remain to be a challenge. This can be explained by several reasons. First, the laser carrier frequency has to be locked to a cavity mode using, for instance, the Pound-Drever-Hall (PDH) technique [2] with a fast feedback loop servo to control high frequency laser fluctuations. Secondly, a second locking servo is required to lock the modulation frequency to a cavity FSR, since the FSR varies when the cavity length is scanned in order to scan the laser frequency over an absorption transition. This has to be implemented by the technique proposed by DeVoe and Brewer [3]. These two fast servo locking systems make the NICE-OHMS detector complex and expensive with limited robustness (see [4] and references therein).

Optical feedback cavity-enhanced absorption spectroscopy (OF-CEAS) [5, 6, 7, 8], wavelength modulation off-axis cavity-enhanced absorption spectroscopy WM-OA-CEAS [9, 10, 11] and cavity ring-down absorption spectroscopy (CRDS) (see [12] and references therein) are among the more robust and simple cavity-enhanced absorption techniques. In order to keep a resonance laser beam injection into a cavity within laser frequency tuning in OF-CEAS, a well-known phenomenon of resonant optical feedback locking of the laser frequency to the optical cavity is used [13]. On the contrary, non-resonance off-axis laser beam injection into the cavity and wavelength modulation for detection of the 1st and 2nd harmonics of the signals on the cavity output are used in WM-OA-CEAS. Measurements of the decay time of the light exiting the cavity have been exploited in CRDS for the extraction of absorption. But in all these techniques shot noise limited absorption sensitivities are difficult to achieve because the photodetection occurs within a bandwidth of 0–1 MHz in OF-CEAS [5], 1–20 kHz in WM-OA-CEAS [9, 10, 11] and ~2 kHz – 1 MHz in CRDS, i.e. in the frequency region where a level of laser flicker noise is predominant and well above a shot noise limit. To our knowledge, the interesting approach of a combination of laser phase modulation at high frequencies, equal to a multiple of a free spectral range (FSR) of a high finesse cavity, with optical feedback locking of the laser to the cavity and further detection of the transmitted optical beam at a high modulation frequency (> 100 MHz), where shot noise predominates, has never been reported.

In this paper, the feasibility of the combination of laser optical feedback locking to a high finesse three mirror V-shaped cavity, laser phase modulation at the cavity FSR frequency and frequency modulation detection has been explored. We present first results of transmittance measurements in an optical setup based upon frequency modulation (FM) optical feedback (OF) cavity-enhanced absorption spectroscopy (FM-OF-CEAS) using a continuous-wave (cw) near-IR distributed feedback (DFB) laser and current modulation at a high frequency equal to the cavity FSR frequency.

**Experimental**

The design of an optical sensor based upon FM-OF-CEAS is shown in Figure 1. A cw single-mode DFB laser diode (LDT 5S515-004, Optilab) emitting at 1529.18 nm was placed into a laser diode mount (TCLDM9, Thorlabs). The free running laser linewidth of 1 – 2 MHz (35 – 40 mA, 2 – 3 mW) was estimated using a technique described elsewhere [14]. The current and the laser temperature (+17 – +20 °C) were controlled by means of a current controller (LDC500, Thorlabs) and a temperature controller (TEC2000, Thorlabs). A voltage-controlled oscillator (ZX95-200S+, Mini-Circuits) was connected to a radio-frequency (RF) power splitter (ZFSC-3-13-S), one output of which was used after attenuation (30 dB) for a 121 MHz modulation of the laser current via a bias-tee of the laser diode mount. A periodic (10 Hz) voltage ramp (1000 points per single scan) was generated by means of a 16-bit data acquisition (DACQ) card (USB-6212BNC, National Instruments) under a LabView program and applied to the analogue modulation input of the current driver for the generation of the laser diode current ramp. An aspheric lens (C330TME-C, Thorlabs) was used for collimation of the horizontally polarised laser diode radiation. A collimated laser beam propagated through a beam splitter (BSF10-C, Thorlabs), an adjustable optical isolator (IO-4-1550-VLP, Thorlabs) and a half-wave plate λ/2 (WPMH05M-1550, Thorlabs). After reflection by a mirror M0 mounted on a piezoelectric transducer PZT-$\phi$ (HPSt 1000/25-15/15VS35, Piezomechanik) the beam was focused by means of a lens (focal length of 50 cm) into a V-shaped cavity formed by three high reflectivity low loss concave mirrors M1, M2 and M3 (curvature radius of −1 m, reflectivity of >99.97 %, Layertec). The laser beam reflected by the cavity mirror M1 was directed into a beam trap. The laser radiation after the half-wave plate was horizontally polarised. It is worth noting that the arrangement of a half-wave plate and a polarising beamsplitter cube typically used for control and adjustment of the optical feedback in OF-CEAS [5] was found to be not sufficient for adjusting the optimal level of optical feedback between the cavity and the laser. The V-shaped cavity had two equal 61.98 cm long arms and a fold-in angle of 9.25°, a free spectral range of 121 MHz, a stability parameter of 0.011 [15] and a beam waist radius of 475 µm in the middle of the each cavity arm. A He-Ne laser and a flip-flop mirror inserted after the half-wave plate were used in an initial alignment of the cavity, whilst a vidicon (K30, Siemens) used for monitoring of final adjustments of the cavity mirrors was placed behind the cavity mirror M3. The V-shaped cavity (Fox-Smith design [16, 17]) was found as the easiest to align and to adjust optimal optical feedback for laser locking to the cavity in comparison against arrangements with a bow-tie cavity or a triangle cavity.

In order to adjust the distance between the laser diode and the cavity mirror M1 close either to 61.98 cm or to 124 cm the laser diode mount was placed on a translation stage. Fine adjustment of the distance and the phase of optical feedback were conducted by varying a voltage applied to the PZT-$\phi$ by means of a single voltage amplifier (SQV1/500, Piezomechanik). Laser radiation transmitted through the cavity mirror M2 was focused by means of an off-axis parabolic mirror OAM1 onto an InGaAs photoreceiver PD1 (Model 1811, Newfocus) with a bandwidth of 0–125 MHz. The optical setup was built by positioning all optical components and the laser mount in open air on an optical table with no vibration isolation. In order to evaluate properties of the frequency-modulated laser beam transmitted through the cavity a 1.5-cm long low pressure reference cell RC1 was inserted between the cavity mirror M2 and the off-axis mirror OAM1. A part of the laser beam, sampled by the beam splitter, propagated through a reference 19-cm long low pressure cell RC2 and was then focused by means of a second off-axis parabolic mirror OAM2 on an InGaAs photoreceiver PD2 (PDA10CF-EC, Thorlabs) with a bandwidth of 0–150 MHz. The DC and AC components of the PD1 and PD2 outputs were separated by means of the bias-tees 1 and 2 (ZX85-12G-S+, Minicircuits), respectively. The AC components of the PD1 and PD2, amplified by means of two low noise amplifiers (ZFL-500LN+, Minicircuits), were then mixed in two mixers MX1 and MX2 (ZX05-1-S+, Minicircuits), respectively, with the reference 121 MHz signals from the power splitter outputs. The phases of the reference frequency signals could be tuned by means of two variable phase shifters PHS1 and PHS2 (JSPHS-150+, Minicircuits) from 0° to ~237°. The two signals from the mixers were filtered by two low pass 10 kHz (−3 dB) filters, respectively, and then recorded simultaneously with the DC components on the hard drive of the laptop for further analysis.

An average mirror reflectivity of 99.97 % was estimated from the measurements of the ring-down time of the V-shaped cavity output (a cavity finesse $F$ of ~10465) and compared to the phase-shift cavity ring-down spectroscopy measurements [18]. The cavity path length enhancement gain of 3333 and an effective absorption path length of 4133 m were obtained. As a test gas we used acetylene ($C_2H_2$) which absorbs around the 1529 nm laser wavelength. The linewidth (half width at half maximum) of $C_2H_2$ is about 2.4 GHz at atmospheric pressure for the $C_2H_2$ absorption line at

1529.18 nm (6539.4537 cm$^{-1}$, C$_2$H$_2$ line strength of 1.144 ×10$^{-20}$ cm$^{-1}$ molecule$^{-1}$ at 296 K [19]). Figure 2 shows typical cavity output intensity and FM signals due to C$_2$H$_2$ absorption in the V-cavity at atmospheric pressure after releasing small amount of acetylene into ambient air close to the V-cavity. In order to observe locking to either even or odd modes [5] only with a frequency spacing of 242 MHz between the cavity output intensity peaks, the distance between the laser diode and the mirror M1 was set to ~124 cm. The laser was locking periodically to the approximately 95 successive either even or odd cavity modes. However, even for very strong absorption FM signal amplitudes observed were very low, whilst stable and reproducible wide scans over 1 cm$^{-1}$ was difficult to achieve with our setup. These low signal-to-noise ratios of FM signals were expected, as it is well known that in order to achieve FM signal detection with maximal signal-to-noise ratio the modulation frequency should be around 1–2 GHz, i.e. beyond the bandwidth of the available photodetector and the maximal frequency of a voltage controlled high frequency oscillator. Thus, in order to estimate best absorption sensitivity that could be achieved for FM signals in our setup, the reference cell RC1 filled with C$_2$H$_2$ at low pressure was introduced for characterisation of the FM light emerging from the cavity and experiencing losses and phase shifts after propagation through the C$_2$H$_2$ absorber, whilst in the following experiments the distance between the laser diode and the cavity mirror M1 was set to 61.98 cm for excitation of successive even and odd cavity modes with frequency spacing of 121 MHz. In order to observe stable and reproducible cavity output signal scans narrow laser frequency tuning over ~0.125 cm$^{-1}$ (31 cavity modes) was used.

**Optical locking of FM light to a V-shaped cavity with absorber**

Under the modulation of the current at a frequency $\nu_{fm}$ the optical field is given by [20, 21, 22, 23]

$$E(t) = A[1 + M\sin(2\pi\nu_{fm}t + \psi)] \times \exp\left(i\left(2\pi\nu_0 t + \beta\sin(2\pi\nu_{fm}t)\right)\right) \quad , \quad (1)$$

where $\nu_0$ is the carrier frequency, $M$ is the amplitude modulation (AM) index, $\beta$ is the frequency modulation (FM) index, and $\Psi$ is the AM-FM phase difference. For small AM and FM indices the optical field can be simply described by a carrier with the frequency $\nu_0$ and two sidebands with the frequencies $\nu_0 \pm \nu_{fm}$ (FM triplet)

$$E(t) = A\exp(i2\pi\nu_0 t) \times \sum_{n=-1}^{n=+1} a_n(\beta, M, \Psi)\exp(i2\pi k\nu_{fm}t) \quad , \quad (2)$$

where

$$a_n(\beta, M, \Psi) = J_n(\beta) + \frac{M}{2}\left[J_{n-1}(\beta)\exp(i\Psi) - J_{n+1}(\beta)\exp(-i\Psi)\right] \qquad (3)$$

and $J_n(\beta)$ is the *n*-th order Bessel function.

For the case of a finite linewidth of the optical field the phase fluctuations $\varphi(t)$ should be included and the equation (2) is given by

$$E(t) = A\exp(i[2\pi\nu_0 t + \varphi(t)]) \times \sum_{n=-1}^{n=+1} a_n(\beta, M, \Psi)\exp(i2\pi k\nu_{fm}t) \qquad . \qquad (4)$$

In the total case of weak optical feedback from the V-shaped cavity, the phase and amplitude of the optical field may differ for each component of the FM triplet. In such case the optical field can be expressed by

$$E(t) = \exp(i2\pi\nu_0 t) \times \sum_{n=-1}^{n=+1} E_n(t)\exp(i[2\pi n\nu_{fm}t + \varphi_n(t)]) \qquad , \qquad (5)$$

where $\varphi_n(t)$ and $E_n(t)$ are the phase variations and the electric field amplitudes for the *n*-th electric field components.

Following the approach introduced in [13], for weak optical feedback from the external V-shaped cavity the variations of the electric field amplitude $E(t)$ at the laser diode with a facet amplitude reflectivity $r_0$ can be derived from the rate equations similar to the one obtained in [24, 25, 26]

$$\begin{aligned}\frac{d}{dt}(E(t)) &= \left[i\omega_N + \frac{1}{2}(G-\Gamma)(1+i\alpha)\right]E(t) + \\ &+ \sum_{n=-1}^{n=+1}\sum_{m=0}^{\infty} K_{nm}E_n(t-\tau_{nm})\exp(i[2\pi(\nu_0 + n\nu_{fm})(t-\tau_{nm}) + \varphi_n(t-\tau_{nm})])\end{aligned} \qquad , \qquad (6)$$

$$\frac{dN(t)}{dt} = J - \frac{N(t)}{\tau_e} - G\left(|E_{-1}(t)|^2 + |E_0(t)|^2 + |E_1(t)|^2\right) \qquad , \qquad (7)$$

where $G$ is the net rate of stimulated emission, $\Gamma$ is the photon decay rate including facet loss, $\alpha$ is the phase-amplitude coupling factor [27], $\tau_{nm} = \tau_{nd} + (2m+1)\tau_{np}$ is the roundtrip delay time between the laser and the cavity mirror M1 for each reflection from the cavity, whilst $\tau_{nd} = 2n_{dc}L_0/c$ is the roundtrip time between the laser output facet and the mirror M1 and $\tau_{np} = 2n_{abs}L_1/c$ is the roundtrip trip time in the arm of the V-shaped cavity, $n_{dc}$ is the refractive index of the medium between the laser and the cavity at the frequency $\nu_0+n\nu_{fm}$, $n_{abs}$ is the refractive index of the gas in the cavity at the frequency $\nu_0+n\nu_{fm}$, $c$ is the speed of light in vacuum, $L_0$ is the distance between the laser diode output facet and the mirror M1, $L_1$ is the length of the cavity arm between the mirrors M1 and M2, $\omega_N = K\pi c/\eta l_d$ is the angular laser diode cavity mode frequency without optical feedback from the cavity, $K$ is an integer number and $\eta l_d$ is the diode laser optical pathlength, $N(t)$ is the carrier density in the laser active layer, $\tau_e$ is the spontaneous lifetime of the excited carriers, $J$ is the injection rate of the excited carriers.

Assuming single pass amplitude absorption $\delta_n$ at the frequency $\nu_0 + n\nu_m$ ($n = -1, 0, +1$) in the cavity $L_1$ long arm, the $K_{nm}$ coefficients can be derived for each FM triplet component with the frequency $\nu_0 + n\nu_m$ similar to the ones obtained by Laurent et al [25] with an additional factor due to absorbance in the V-shaped cavity

$$K_{nm} = \frac{c}{2\eta l_d} \sqrt{\beta_P} \frac{1-r_0^2}{r_0} (1-r^2) r \exp(-2\delta_n) \left(r^4 \exp(-4\delta_n)\right)^m, \qquad (8)$$

where $\beta_p$ is the power mode coupling factor and $r$ is the amplitude reflectivity of the cavity lossless mirrors.

After separation the FM triplet harmonics the expression (6) gives a set of six equations for the phases and the amplitudes of each electric field component. After taking the time-independent solutions for $E_n(t)$ and $\varphi_n(t)$ the real and imaginary parts of these equations can be separated for the carrier ($n=0$) and the two sidebands ($n = -1, +1$) giving the equations for each FM triplet component

$$2\pi(\nu_0 + n\nu_{fm}) = \omega_N + \frac{\alpha}{2}(G - \Gamma) - \sum_{m=0}^{\infty} K_{nm} \sin(2\pi(\nu_0 + n\nu_{fm})\tau_{nm}) \quad, \qquad (9)$$

$$G - \Gamma = -2 \sum_{m=0}^{\infty} K_{nm} \cos(2\pi(\nu_0 + n\nu_{fm})\tau_{nm}) \quad . \qquad (10)$$

After substitution of $(G-\Gamma)$ from (10) in (9) the laser frequencies of the FM triplet components in case of optical feedback from the V-shaped cavity can be expressed as

$$2\pi(\nu_0 + n\nu_{fm}) = \omega_N - (1+\alpha^2)^{1/2} \sum_{m=0}^{\infty} K_{nm} \sin(2\pi(\nu_0 + n\nu_{fm})\tau_{nm} + \theta) \;, \qquad (11)$$

where $\theta = \arctan(\alpha)$. By substituting the $K_{nm}$ coefficients from (8) in (11) the laser frequencies of the FM triplet are given by

$$2\pi(\nu_0 + n\nu_{fm}) = \omega_N - K_n \frac{\sin\left[2\pi(\nu_0 + n\nu_{fm})(\tau_{nd} + \tau_{np}) + \theta\right]}{1 + \left[F_n \sin(2\pi(\nu_0 + n\nu_{fm})\tau_{np})\right]^2} +$$
$$+ K_n \frac{r^4 \exp(-4\delta_n) \sin\left[2\pi(\nu_0 + n\nu_{fm})(\tau_{nd} - \tau_{np}) + \theta\right]}{1 + \left[F_n \sin(2\pi(\nu_0 + n\nu_{fm})\tau_{np})\right]^2} \;, \qquad (12)$$

where

$$F_n = \frac{2r^2 \exp(-2\delta_n)}{1 - r^4 \exp(-4\delta_n)}$$

and

$$K_n = (1+\alpha^2)^{1/2} \frac{c}{2\eta l_d} \sqrt{\beta_p} \frac{1-r_0^2}{r_0} r \exp(-2\delta_n) \frac{1-r^2}{\left(1-r^4 \exp(-4\delta_n)\right)^2} \quad .$$

The derived equation (12) can be simplified to the equation (8) in [25] and to the equation (3) in [5] for the case of no absorption neither in the cavity nor between the laser and the cavity. From the equation (12) it follows that the frequencies of the carrier and the two sidebands display a behaviour which is qualitatively similar to the behaviour of the laser frequency of a single-mode laser coupled via optical feedback either to a tilted confocal Fabry-Perot cavity [25] or a V-shaped cavity [5].

**Results and discussion**

The transfer function of the V-shaped cavity for the beam passed through the cavity mirror M2 is given by

$$V(\nu) = \frac{t^2 r^2 \exp(-3(\delta + i\phi))}{1 - r^4 \exp(-4(\delta + i\phi))} \quad , \tag{13}$$

where $t$ is the amplitude transmittance of the mirror, $\delta$ and $\phi$ are the amplitude attenuation and phase shift experienced by optical field at the frequency $\nu$ after single pass through one $L_1$ long cavity arm. The amplitude attenuation and the phase shift can be derived in analytical form for Lorentzian [28], Doppler broadened [29] and Voigt [4] line profiles. The electric field of the transmitted light is given by

$$E_T(t) = \sum_{n=-1}^{n=1} V(\nu_0 + n\nu_{fm}) E_n(t) \exp\left(i\left[2\pi(\nu_0 + n\nu_{fm})t + \varphi_n(t)\right]\right) \quad . \tag{14}$$

The power of the transmitted beam measured by the photodetector PD1 is $P_{PD1}(t) = |E_T(t)|^2$ and after some algebraic transformations [30] the power becomes

$$P_{PD1}(t) = \sum_{n=-1}^{n=1} P_n(t)|V(\nu_0 + n\nu_n)|^2 +$$

$$+ \cos(2\pi\nu_{fm}t)\,\mathrm{Re}\left\{\begin{array}{l} E_0E_{-1}e^{i(\varphi_0-\varphi_{-1})}V(\nu_0)V^*(\nu_0-\nu_{fm}) + E_0E_{-1}e^{i(\varphi_{-1}-\varphi_0)}V(\nu_0-\nu_{fm})V^*(\nu_0) + \\ + E_0E_1e^{i(\varphi_1-\varphi_0)}V(\nu_0+\nu_{fm})V^*(\nu_0) + E_0E_1e^{i(\varphi_0-\varphi_1)}V(\nu_0)V^*(\nu_0+\nu_{fm}) \end{array}\right\} +$$

$$+ \sin(2\pi\nu_{fm}t)\,\mathrm{Im}\left\{\begin{array}{l} E_0E_{-1}e^{i(\varphi_{-1}-\varphi_0)}V(\nu_0-\nu_{fm})V^*(\nu_0) - E_0E_{-1}e^{i(\varphi_0-\varphi_{-1})}V(\nu_0)V^*(\nu_0-\nu_{fm}) - \\ - E_0E_1e^{i(\varphi_1-\varphi_0)}V(\nu_0+\nu_{fm})V^*(\nu_0) + E_0E_1e^{i(\varphi_0-\varphi_1)}V(\nu_0)V^*(\nu_0+\nu_{fm}) \end{array}\right\} +$$

$$+ (2\nu_{fm}\,\mathrm{terms}) \quad .(15)$$

At the phase modulation frequency $\nu_{fm}$ close to the cavity FSR frequency the FM triplet propagates through the cavity resulting in the DC and AC terms. The DC term in equation (15) represents the sum of the intensities of the FM triplets attenuated due to absorption in the cavity. The terms at the frequency $\nu_{fm}$ arise from the interference between the carrier and sidebands, whilst the $2\nu_{fm}$ terms originate from the sideband interference with each other.

The FM index $\beta$ of 0.38±0.03 the AM index $M$ of 0.007± 0.003 and the phase shift $\Psi$ of 3.14±0.2 were estimated from the absorption/dispersion signals recorded for the case with no optical feedback from the cavity and the pressure of 856 mB in the reference cell RC2 filled with $C_2H_2$ diluted in air. The laser FM parameters were extracted from the fits of the experimental MX2 output FM signals by a sum of an in-phase component and a quadrature component [31] derived for a Lorentzian lineshape, whilst the experimental FM signals were recorded at various phases of the phase shifter PHS2.

In order to propagate effectively through the cavity, the modulation frequency $\nu_{fm}$ of the laser light should be adjusted close to the cavity FSR frequency, whilst the linewidths of the FM triplet components should be close or within the cavity bandwidth of 11.562 kHz. The modulation frequency $\nu_{fm}$ was tuned by varying a voltage set of ~5 V on the input of the VCO. Narrowing of the laser linewidth was achieved by adjusting the level of optical feedback ($10^{-5} - 10^{-4}$) by means of a rotation the output polariser of the optical isolator. Stable and reproducible successive odd and even modes were observed within tuning the laser frequency. Simulations of the FM signals described by the equation (15) show that the error signal can be detected from the $\nu_m$ term of the *sine* function within the deviation of the modulation frequency from the FSR frequency. This error signal can be used in a digital proportional-integral-differential (PID) control for locking the modulation frequency $\nu_{fm}$ to the cavity FSR. The experimental confirmation for the extraction of such error signal from the transmitted FM signal of the V-shaped cavity is shown in Fig. 3. The DC outputs of the detector PD1 and PD2 with the filtered outputs of the mixer MX1 and MX2 were recorded for different modulation frequencies $\nu_{fm}$ tuned around the cavity FSR. The positive and negative slopes of the FM signals within locking of the laser to the cavity modes were observed for $\nu_{fm} > \nu_{FSR}$ (Fig. 3a) and $\nu_{fm} < \nu_{FSR}$ (Fig. 3b), whilst for the modulation frequency close to the FSR frequency the FM signal was flattened (Fig. 3c). The transmittance signals shown in the Fig. 3 were recorded for a distance between the cavity and the laser that was adjusted by a few mm away from the optimal distance of ~62 cm. For such a case the shapes of the transmitted intensity peaks vary markedly within a single scan from typical symmetric rounded peaks observed at the optimal optical feedback phase. However, even for asymmetric transmitted intensity peaks with negative and positive slopes the error signal of the deviation of the modulation frequency from the cavity FSR frequency was close to the error signal observed for the optimal optical feedback phase with symmetric and rounded transmitted intensity peaks. No FM signals were observed when the modulation frequency was tuned by a few tens kHz away from the FSR frequency of 121 MHz. The transmitted intensity (trace 3 in the Fig. 3) of the long reference optical cell RC2 filled by $C_2H_2$ and the FM signal extracted from the photodetector PD2 output were used for visual observation of both the laser frequency locking to the successive cavity modes and then the laser frequency jumps from one cavity mode to the proceeding adjacent one.

It is well known in FM spectroscopy [31] that the in-phase component of the FM signal is proportional to the absorption and the quadrature component is proportional to the dispersion induced by the absorption line. In order to estimate the baseline noise of the FM triplet intensity and the FM signal the short reference cell RC1 was filled by $C_2H_2$ at a pressure of 40 mB. The distance between the laser and the cavity was adjusted to ~62 cm, whilst fine tuning of the feedback phase was completed via precise positioning of the mirror M0 by means of the PZT-$\phi$. The frequency of the VCO was tuned close to the cavity FSR frequency by observation of the flattening of the slope of the FM signal within the optical locking of the laser to the cavity modes. Fig. 4 shows the transmitted FM triplet raw intensity and the filtered mixer MX1 output for an averaging of ten 100-ms successive scans. The DC and FM signals from the photodetector PD2 are also shown. Fast fluctuations in the intensity and the FM signal were smoothed by applying to the acquired intensity and FM signals a Fast Fourier Transform (FFT) filter with the noise-equivalent bandwidths of 4 kHz and 2 kHz, respectively. The FFT filtered signals are shown in Fig. 5. It is worth noting that the FFT filter bandwidths were selected to be minimal under condition of minimal distortion of the signals around the points of the intensity maxima.

For the transmitted intensity signal (trace 1 in Fig. 5) the maxima points were found for each intensity peak, when the laser was locking to the cavity. The processed transmitted intensity spectrum and the FM signal at the points of the intensity maxima are shown in Fig. 6. The maxima in the processed intensity signal vary for odd and even modes (see explanation in [5]), whilst variations of the FM signal at the points of the transmitted intensity maxima are markedly smaller.

Fig. 7 shows the absorption spectra calculated from the selected intensity peak points with the maximal amplitude. The maximal intensity points were fitted by a second order polynomial for a baseline and a Voigt profile function for the $C_2H_2$ absorption line. Fig. 7 shows the absorption experimental points calculated as a natural logarithm of the ratio of the fitted baseline value to the intensity at the point of maximal intensity and a Voigt lineshape fit. For the Voigt fit an approximation by a weighted-sum of Gaussian and Lorentzian functions with a full width at half maximum of 651(8) MHz and a weighted factor of 0.69(5) was used [32]. A conversion factor of 121 MHz was used for the conversion of time axis in Fig. 5 into the relative frequency axis in Fig. 6. For 10 averaged and subsequently FFT filtered measurements over the 16 cavity modes (either even or odd) an absorbance baseline noise (standard deviation) of 0.0039 was estimated for the laser frequency tuning across the $C_2H_2$ absorption line, thus resulting in a signal-to-noise ratio of 168. For the effective beam path length of 4133 m in the V-shaped cavity such a baseline noise of the absorption spectra provides an absorption sensitivity of $9.4\times10^{-9}$ cm$^{-1}$. The observed signal-to-noise ratios of the 10 averaged and then FFT filtered FM components (see Fig. 8) were typically by a factor of ~10 (for some recorded spectra up to 20) higher. Thus we assume that a projected absorption sensitivity of $9.4\times10^{-10}$ cm$^{-1}$ can be achieved for absorption measurements extracted from the FM signals. The signal-to-noise ratio for the FM signal in FM-OF-CEAS can be estimated from the ratio of a FM peak-to-peak amplitude either to the standard deviation of the residual between FM experimental points and a FM signal fit or to the standard deviation of the absorbance baseline with no $C_2H_2$ absorber in the short optical cell. In the first approach, the common working equation in FM spectroscopy (see eq. 3 in [28]) for a weighted sum of absorption and phase components derived for Gaussian [29] and Lorentzian [28] functions was used in the fit of the FM signal in the Fig. 8 with the fitting parameters for the FM index of 0.38, the AM index of 0.007, the phase shift of 3.14, the Voigt fit weighted factor of 0.65 [32] and the absorption line FWHM of 648 MHz. In the second approach, the 1.5-cm long optical cell was filled with nitrogen and the FM baseline noise was evaluated for the laser locked to the V-cavity and the FM triplet passing through the V-cavity. The signal-to-noise ratios estimated within both approaches were comparable.

The theoretical shot-noise limited FM absorption sensitivity normalised by a noise-equivalent bandwidth in FM-OF-CEAS with a V-shaped cavity formed by two $L_1$ long arms can be derived similar to equations (2) and (3) obtained in [1] and is given by

$$\alpha_{min} = \frac{\pi}{F 2 L_1} \left( \frac{2e}{S_{PD} P_0} \right)^{1/2} \frac{\sqrt{2}}{J_0(\beta) J_1(\beta)} \quad , \quad (16)$$

where $e$ is the elementary charge, $S_{PD}$ is the responsivity of the photodetector, $P_0$ is the power incident on the photodetector PD1 and a factor $F2L_1/\pi$ corresponds to an effective cavity path length [12]. For the measured transmitted power of 10 µW, the maximum possible value of $J_0(\beta)J_1(\beta)$ of 0.34 and $S_{PD}$=1 A/W, one computes a shot-noise limited noise-equivalent absorption sensitivity of $1.8\times10^{-12}$ cm$^{-1}$Hz$^{-1/2}$. Here, the noise-equivalent absorption sensitivity of $2.1\times10^{-11}$ cm$^{-1}$Hz$^{-1/2}$ was estimated for the typical FM spectra recorded with our FM-OF-CEAS setup, processed and then smoothed by means of the FFT filter with the noise-equivalent bandwidth of 2 kHz. Thus, the absorption sensitivity projected for the measurements in the FM-OF-CEAS setup is a factor of 11.7 (5.8 for some recorded FM signals) above the shot-noise limit.

The projected noise-equivalent absorption sensitivity of $2.1\times10^{-11}$ cm$^{-1}$ Hz$^{-1/2}$ corresponds to a minimum detectable $C_2H_2$ number density of $\sim5.2\times10^7$ molecule cm$^{-3}$ or ~20 ppt (parts per trillion) at a pressure of 100 mB. Further improvement of the absorption sensitivity is expected for faster scan rates, higher cavity output intensity and, finally, higher reflectivity and lower losses of the cavity mirrors. The absorption sensitivities close to shot noise limits in FM-OF-CEAS may lead to the detection of even smaller number densities of molecules down to tens of ppq (part per quadrillion) for strongly absorbing $CO_2$ and $N_2O$ at mid-IR wavelengths. The detection of $C_2H_2$, $H_2O$, HF, $NH_3$ at ppt level may find applications in nanotechnology for monitoring the purity of components in the production of 3D nanostructures and production of high purity gases. Monitoring radio-labelled $^{11}CO_2$, $^{11}CO$, $H^{11}CN$, $^{11}CH_4$ molecules at ppt concentrations and in small volumes of ~10 mL [33] using FM-OF-CEAS may find application in the optimisation of the production of radio-labelled minute quantities of drugs for positron emission tomography (PET) [34].

**Conclusions**

We have characterised the potential of frequency modulation and optical feedback locking for cavity-enhanced absorption spectroscopy with a near-IR DFB laser at 1529.18 nm and a V-shaped cavity. The noise-equivalent absorption sensitivity of $2.1\times10^{-11}$ cm$^{-11}$Hz$^{-1/2}$ was estimated for the typical FM spectra recorded with our setup. This was a factor of 11.7 above the shot-noise limit. Further improvement of the absorption sensitivity is expected for higher cavity output power, higher mirror reflectivity and implementation of two digital PID controllers for adjusting of the optimal optical phase and locking the laser modulation frequency to the cavity FSR frequency. The full potential of FM and OF for CEAS based upon near-IR and mid-IR lasers with this high shot-noise limited absorption sensitivity is yet to be realised. Successful implementation of the proposed improvements for the FM-OF-CEAS may enable absorption measurements close to shot-noise limited sensitivities.

**Acknowledgments**

The research was supported by the European Commission within a Marie-Curie experienced researcher Intra-European Fellowship Award to V. Kasyutich (Kasiutsich). We also acknowledge the support by ETH Zurich and are very grateful to K.M.C. Hans, J. Kottmann, J.M. Rey and M. Gianella for their help and support during this project.

**Figure captions.**

Figure 1. Setup for frequency modulation optical feedback cavity-enhanced absorption spectroscopy in a V-shaped cavity formed by three high reflectivity mirrors M1, M2 and M3: PD1 and PD2 are the photodetectors; $\lambda/2$ is the half-wavelength plate; OAM1 and OAM2 are the off-axis parabolic mirrors; RC1 is the 1.5-cm long low pressure reference cell; RC2 is the 19-cm long low pressure reference cell; MX1 and MX2 are the RF mixers; PHS1 and PHS2 are the phase shifters; 121 MHz VCO is the voltage-controlled oscillator at the central frequency of 121 MHz, DACQ card is the 16-bit data acquisition card.

Figure 2. The intensity signal (trace 1) of the V-shaped cavity and the FM signal (trace 2) recorded for $C_2H_2$ (~8 ppm) in ambient air at atmospheric pressure of 1013 mBar after releasing $C_2H_2$ near the V-cavity. The distance between the laser diode and the cavity mirror was ~124 cm, whilst frequency spacing between the cavity mode peaks was 242 MHz.

Figure 3. The intensity signal (trace 1) of the V-shaped cavity and the reference cell RC1 for $\nu_{fm} > \nu_{FSR}$ (Fig. 2a), $\nu_{fm} < \nu_{FSR}$ (Fig. 2b) and $\nu_{fm} = \nu_{FSR}$ (Fig. 2c). The positive (Fig. 2a), negative (Fig.2b) and zero (Fig.2c) slopes are observed for the FM signal on the PD1 output (trace 2) within locking the laser frequency to the cavity resonance modes. The intensity signal (trace 3) and the FM signal (trace 4) are for the reference cell RC2 recorded at the $C_2H_2$ pressure of 35 mB. For clarity the traces 3 and 4 are shifted down by 0.15 V (a), 0.2 V (b) and 0.2 V (c).

Figure 4. The transmitted intensity (trace 1) and FM (trace 2) signals of the V-shaped cavity and the reference cell RC1 with $C_2H_2$ for the case of $\nu_{fm} = \nu_{FSR}$. The transmitted intensity (trace 3) and the FM (trace 4) signals, both shifted for clarity down by 0.2 V, are for the reference cell RC2 recorded at the $C_2H_2$ pressure of 40 mB.

Figure 5. The transmitted FFT filtered intensity (trace 1) and FM (trace 2) signals of the V-shaped cavity and the reference cell RC1 with $C_2H_2$ for the case of $\nu_{fm} = \nu_{FSR}$.

Figure 6. Processed transmitted intensity and FM $C_2H_2$ absorption spectra.

Figure 7. $C_2H_2$ absorption spectra (●), a Voigt fit (red solid line) and a residual (-□-) multiplied by a factor of 10 and shifted down by 0.15 V.

Figure 8. $C_2H_2$ FM absorption spectra (●), a fit (red solid line) and a residual (-□-) multiplied by a factor of 100 and shifted down by 0.15 V.

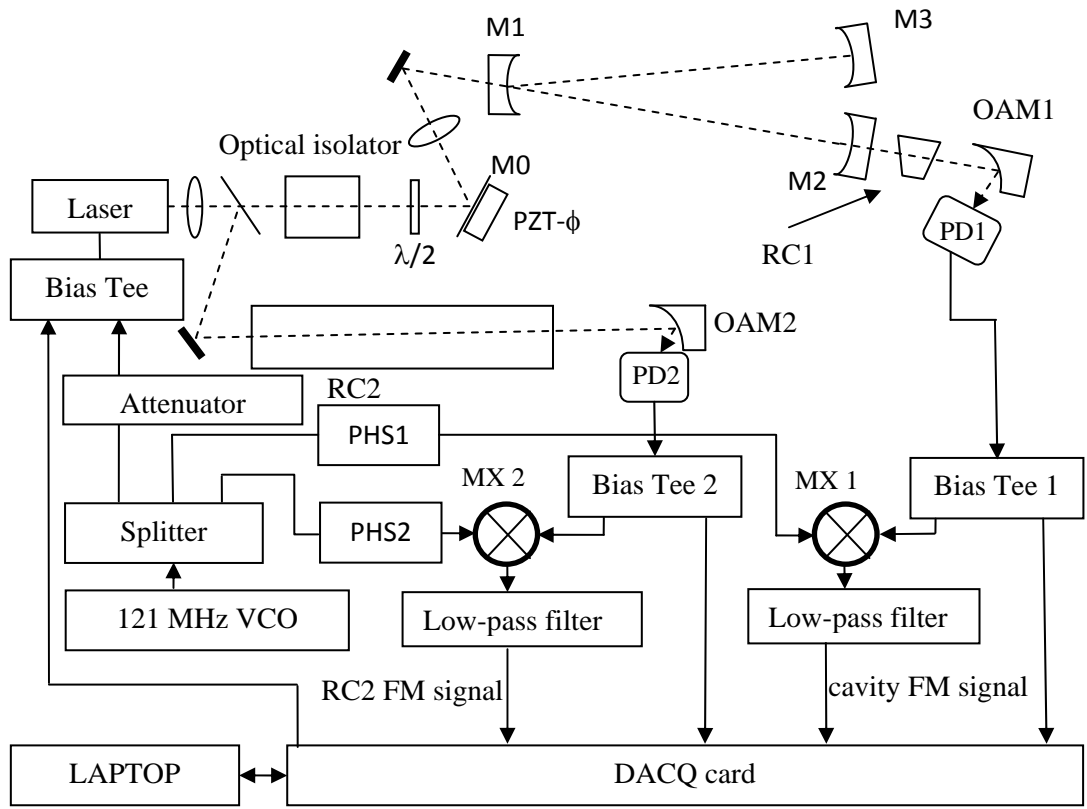

Fig. 1.

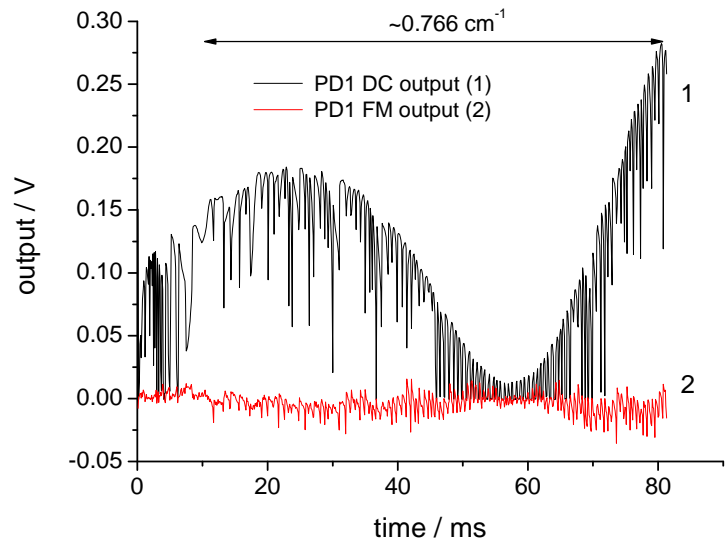

Fig. 2.

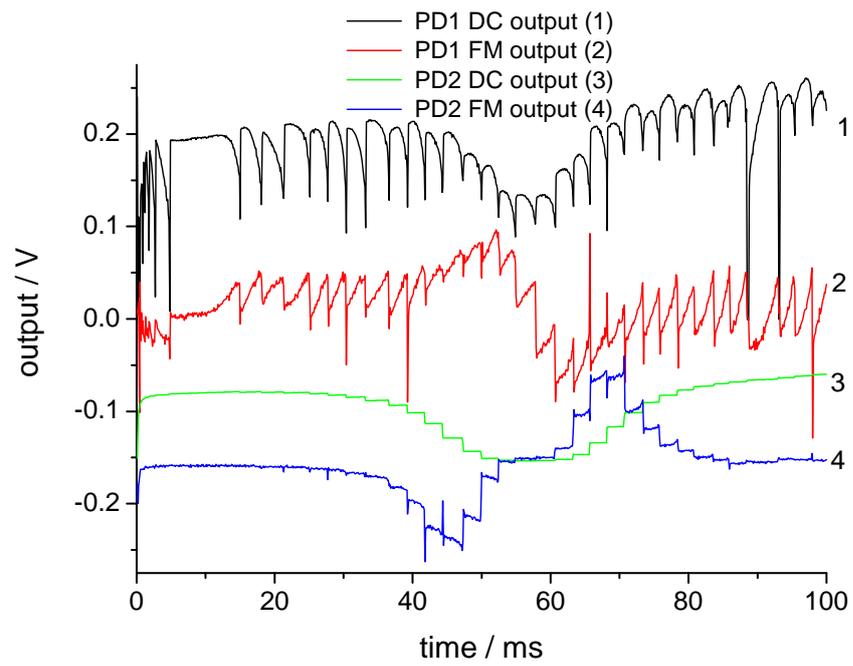

Fig. 3a

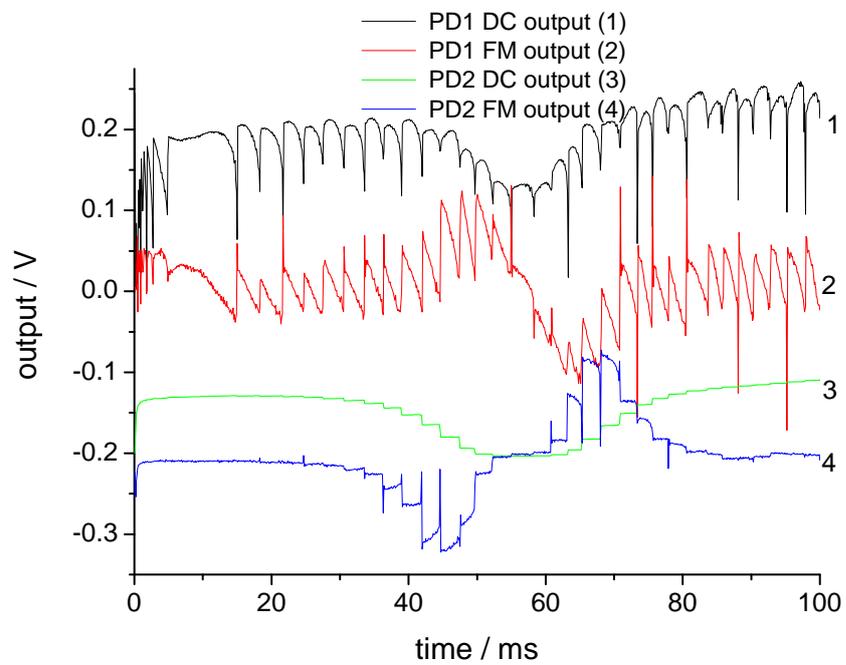

Fig. 3b

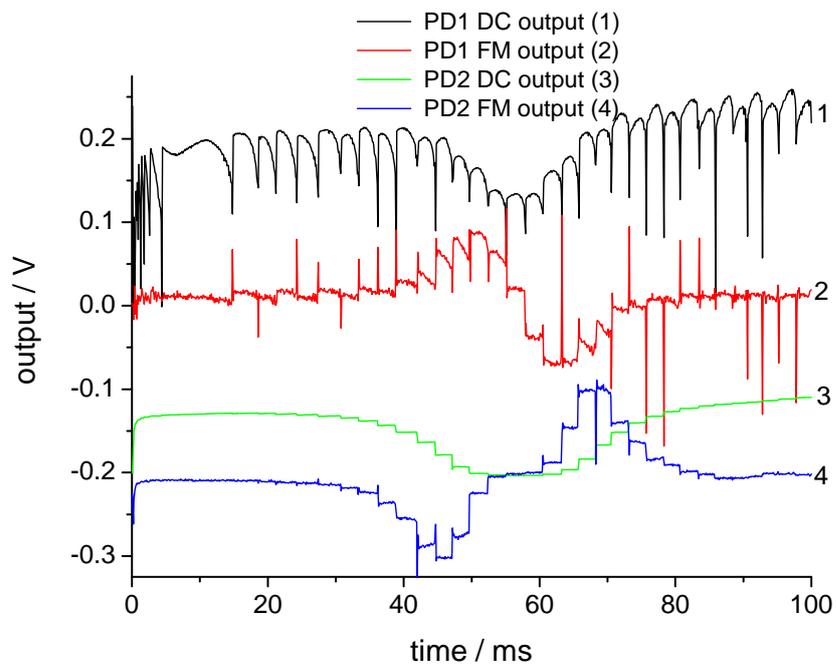

Fig. 3c

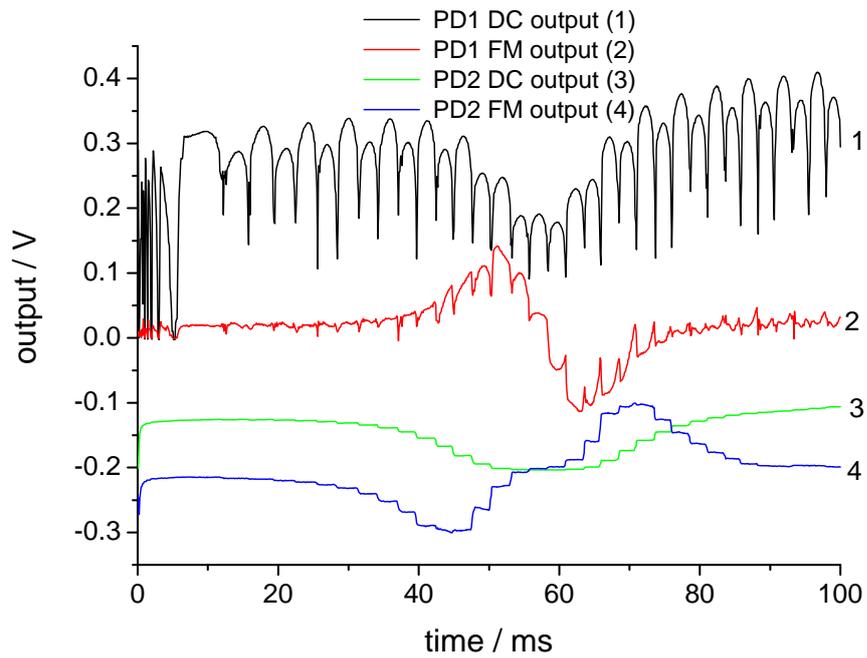

Fig. 4.

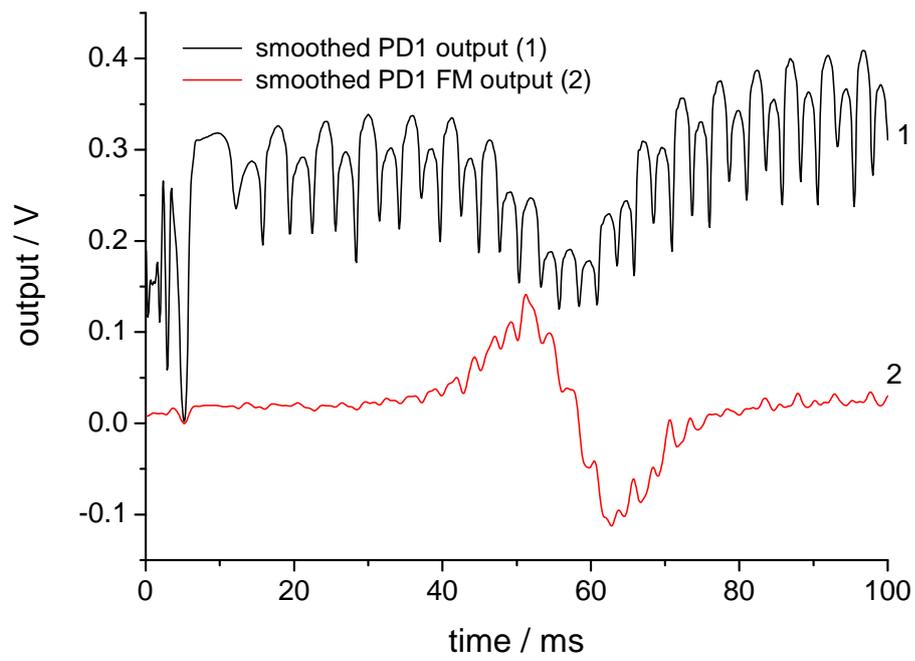

Fig. 5.

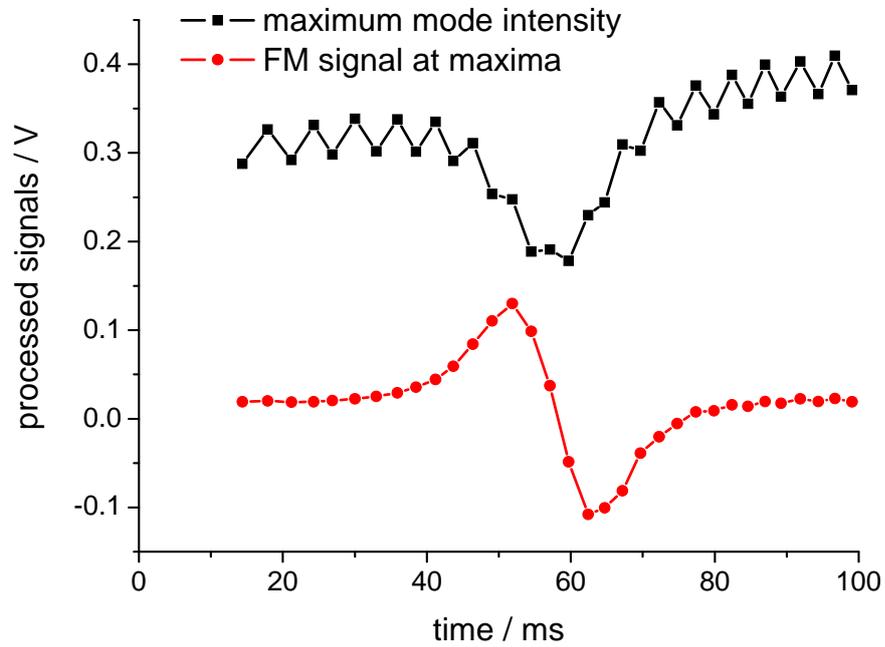

Fig. 6.

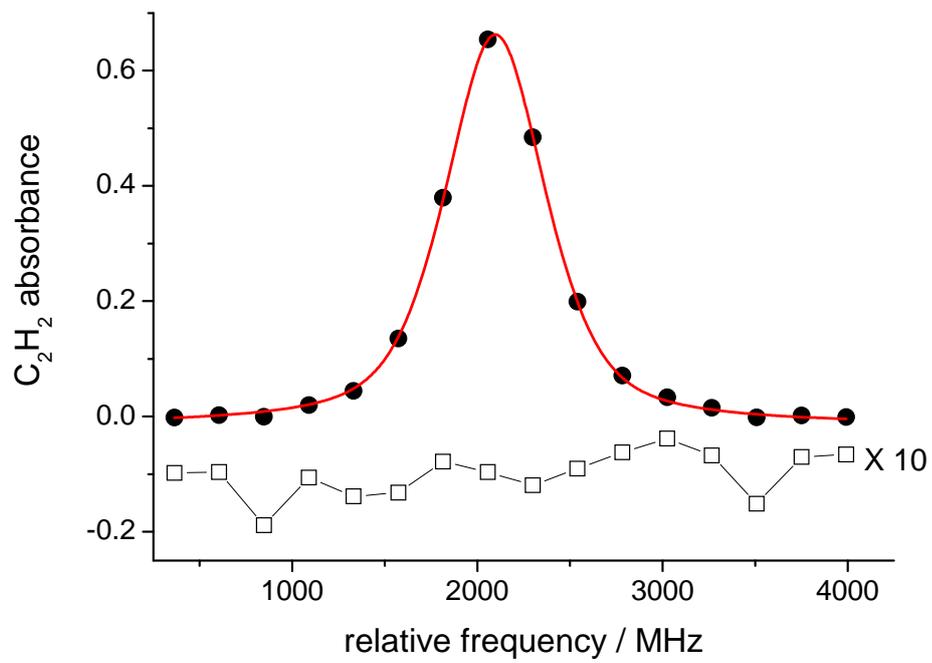

Fig. 7.

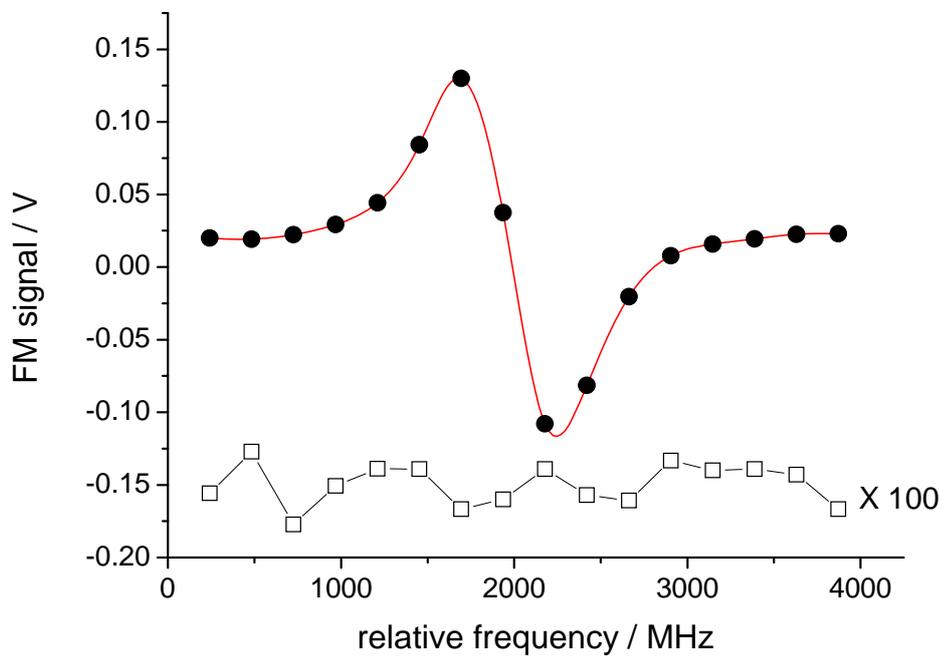

Fig.8.